\documentclass{birkmult}

\usepackage{amsthm,amsmath,amssymb}



\begin{document}

\title[Automated Reasoning in Social Choice]
 {Automated Reasoning in Social Choice Theory -- Some Remarks\textsuperscript{ $\star$}}

\author[S. Chatterjee]{\textbf{Siddharth Chatterjee}}
\address{Indian Statistical Institute, Delhi Center, New Delhi, India}
\email{123sidch@gmail.com}

\thanks{\textsuperscript{$\star$} We thank three referees for their insightful comments.\\}

\author[A. Sen]{\textbf{Arunava Sen}}
\address{Indian Statistical Institute, Delhi Center, New Delhi, India}
\email{asen@isid.ac.in}

\begin{abstract}
  Our objective in this note is to comment briefly on the newly
  emerging literature on computer-aided proofs in Social Choice
  Theory. We shall specifically comment on \cite{Tang} (henceforth TL)
  and \cite{GeistP} (henceforth GE). In the Appendix we provide
  statements and brief descriptions of the results discussed in this
  note.
\end{abstract}

\maketitle

Both of the papers on computer-aided proofs in Social Choice Theory that we have reviewed, TL~\cite{Tang} and GE~\cite{GeistP}, adopt the same approach. They prove impossibility results by reducing a problem of an arbitrary size to a ``small'' base case which is then tackled computationally. TL looks at various versions of the classical Arrovian aggregation problem. It uses conventional induction methods to show that the existence of non-trivial solutions in the general case implies the existence of non-trivial solutions in the base case. It then shows that non-trivial solutions do not exist in the base case thereby establishing an impossibility result. A naive approach to the base case verification is computationally intractable.\footnote {The smallest interesting case is one with two voters and three alternatives. The number of Arrovian aggregators is $36^6$ in this case. A naive approach would be to list each possibility and verify the Independence of Irrelevant Alternative axiom - this is clearly intractable computationally.} The idea in the paper is to encode the properties of an Arrovian Social Welfare Function as 
a Constraint Satisfaction Problem (CSP), which can then be solved using standard algorithmic techniques from computer science. Instead of checking every possible rule for consistency with the axioms, there is an algorithm that iteratively generates rules. If at any point the construction in progress meets a contradiction with any of the axioms applicable in the present stage, the present step and its future course is abandoned. Then a step is backtracked and the algorithm explores other potential future stages of the present construction. This depth-first search based technique eliminates redundancies inherent in the naive approach and turns out to be efficient for practical use in the base case verification.\footnote{The search can be completed with this procedure using a standard processor within a second.} The paper uses this approach to provide proofs of the Arrow, Muller-Satterthwaite and Sen (``Impossibility of a Paretian Liberal'') Theorems in addition to a completely new result.

	GE builds on TL and considers the axiomatic set-ranking problem. It proves a result called the Preservation Theorem that allows the reduction from the general case to the base case for a wide class of axiom systems. It addresses the base case using computational strategies inspired by TL and proves several ($84$) results including the Kanai-Peleg Theorem and some new ones. The new results are generated by considering combinations of standard axioms in this framework.

	We will first comment on the significance of the results in these two papers for social choice theory and then on the general applicability and limitations of the approach.

	The new proofs of existing results are not the most interesting aspects of the papers. Several proofs already exist - some in fact, are based on induction and are more insightful than computational proofs. However, some of the new results are quite striking. Consider, for instance Theorem 5 in TL. The standard proof of Arrow's Impossibility Theorem (\cite{Sen}) using the Field Expansion and Group Contraction Lemmas makes intensive use of the Weak Pareto (WP) axiom. Wilson's Theorem therefore comes as a surprise because it shows that replacing the WP axiom by a substantially weaker range condition additionally allows only inverse dictatorship and the constant rule. The proof of Wilson's theorem  typically proceeds by showing that the range condition in conjunction with the Independence of Irrelevant Alternatives (IIA) axiom implies that the rule is either constant, satisfies WP or an inverse WP axiom (\cite{Malawski}). If WP is satisfied, then dictatorship follows from Arrow's Theorem; if inverse WP holds, 
then an ``inverse'' version of all the arguments in Arrow's Theorem can be replicated to yield an inverse dictatorship theorem. It is clear from the standard proofs that IIA is critical but the role of the additional axiom (i.e. the range conditions) is unclear. In fact, one may conjecture that it is important in order to generate a negative result. Theorem 5 makes it clear that this is not the case. Among the $6^{36}$ rules, only $94$ satisfy IIA. The rules other than dictatorship and inverse dictatorship are not satisfactory. Specifically, restrictions of the value of the rule to alternative triples have a range of at most two and differ from each other by a Kendall distance\footnote{See the Appendix for a definition.} of at most one. As far as we know, the TL result is the only Arrow-type result in the literature that does not use an axiom other than IIA. It clearly demonstrates the powerful role of IIA in reducing possibility results. We believe that this result could not have been conjectured without 
computational aids.

	A similar comment applies to the new results in GE. The axioms considered are familiar and well-motivated - their mutual compatibility is an important issue to be resolved. As pointed out by the authors some of the results have escaped the attention of social choice theorists. For example, \cite{Bossert} claimed to have characterized max-min rules by a set of four axioms  while this paper shows that these axioms lead to an impossibility.\footnote{\cite{Arlegi} had first pointed out that the claim of \cite{Bossert} was incorrect. He had provided an alternative charcterization without showing that the \cite{Bossert} axioms are inconsistent.} Finally, the paper provides direct (or manual) proofs of some of the new results it discovers.

	We believe that automated reasoning can be highly fruitful in addressing a range of problems in social choice theory. Such reasoning applied to a ``small'' base case can be used to verify and generate conjectures. Manually exploring possibilities even for such cases is impossible due to the astronomical number of possibilities. The overall utility of this approach, in our opinion, depends critically on ``how successfully'' the general problem can be reduced to a manageable base case. We note that even if a general reduction in the problem is not feasible, working out examples with small numbers computationally, is valuable for getting insights into the general problem.

	What are the features of the problem that permit a reduction of the desired variety? It may be tempting to infer from TL and GE that only impossibility results are amenable to such reductions. This may be misleading. Virtually all results in social choice theory are of the following sort: a rule satisfies a certain set of axioms if and only if it belongs to a certain class. Any such (characterization) result can be easily reformulated as an impossibility result by asking for the possibility of a rule belonging to the complement of the characterized class yet satisfying the same set of axioms. \footnote{A referee has pointed out an additional issue that may arise even if reduction to a base case is feasible - it may not be possible to cast the base case verification as a CSP due to lack of finiteness. For instance, consider the problem of characterizing the Borda social welfare function. The base case verification of impossibility would involve checking the existence of certain numbers which cannot be cast 
as a CSP. See also footnote 11.} We believe that the key to the reduction issue is therefore, the ``complexity'' of the class of rules under consideration. We illustrate this point with a few examples.

\begin{enumerate}
\item	Dictatorship-type results are relatively easy to address by this method. It is natural to ask, for instance, whether a computer-aided proof for the Gibbard-Satterthwaite Theorem (GST) in strategic voting theory, is possible. In fact this question has been answered in the affirmative in \cite{TangM}.\footnote{See \cite{Schmeidler} for a direct approach along identical lines.} There are, however, a host of open questions regarding dictatorial rules where this approach is likely to be useful. We briefly outline one such class of problems, that of characterizing dictatorial domains.
	According to GST, every strategy-proof and unanimous social choice function defined over the complete domains of strict orderings, is dictatorial.\footnote{This is subject to the condition that there are at least three alternatives.} A domain of orderings (a non-empty subset of the complete domain) is dictatorial if every strategy-proof and unanimous social choice function defined over this domain, is dictatorial. The complete domain is of course, dictatorial but are there other dictatorial domains? Recent results (\cite{ACS}) have shown that the class of dictatorial domains is extremely large - these domains can be very sparse (linear in the number of alternatives).
\footnote{In this respect, the strategic and aggregative problems in social choice are very different from each other.} 
A full characterization of dictatorial domains is not yet in sight. We believe that a computational approach using TL methods will be helpful in this regard. For instance, we have verified, using a SAT solver that had been implemented in SWI-Prolog, that the only dictatorial domain in the case of three alternatives is the complete domain.
\footnote{This result follows from \cite{ACS}. The complexity of the problem is considerably higher for even slightly larger problems.} 
These methods remain feasible for slightly larger problem sizes. There is one aspect of the dictatorial-domains problem that makes this approach promising - the reduction from the general case to the case of two voters is well established (\cite{ArunavaEL}). Unfortunately, the reduction to three alternatives is not valid as the previous comments show. A more sophisticated approach to the reduction of number of alternatives is therefore required. Allowing for indifferences in individual preferences substantially complicates the analysis of dictatorial domains (\cite{Sato}).
\footnote{Indifferences increase the size of the domain. However, size is not the primary complication introduced by indifferences. When a voter misreports her preferences, the outcome can change to a new outcome that is indifferent to the original outcome according to {\it both} original and misreported preferences. This possibility cannot occur with strict preferences. As a result, the analysis with indifference is considerably more complicated.} This is another model where computational verification will be invaluable.

\item	Several important results in social choice theory are characterizations of ``well-behaved'' rules over restricted domains. Perhaps the most prominent of these are the median-voter (type) rules over single-peaked domains. A fundamental result was proved in \cite{Moulin} and a large literature has developed on the subject. Alternatives in this model are a finite number of points on the real line. A single-peaked preference has a unique peak and preferences decline in both directions away from the peak. A median rule picks the median peak among the set of voter peaks and (possibly) some fixed or phantom peaks. Heuristically the median voter rule is a more complex rule than a dictatorship and proving a reduction, especially on the number of alternatives is a more challenging task.

	Another restricted domain problem that admits ``well-behaved'' rules is the classical allocation problem with selfish preferences. There are $n$ voters who have to be allocated at most one of $m$ distinct objects ($m \geq n$). Each voter has a strict ordering over the $m$ objects and is indifferent among all allocations where she is assigned the same object. A rich class of rules are strategy-proof in this model (including the well-known ``top-trading'' cycle) and a characterization exists (\cite{Papai}). Nevertheless, a tighter characterization with fewer axioms is desirable. A reduction argument is  difficult in this case as well. Again, computational exercises for small cases may help in forming conjectures.
\item	There are  several important classical social choice problems where computational approaches are unlikely to be of use because they inherently lack finiteness. These include problems involving randomization (\cite{GibbardR}), cardinalities (ranking of sets based on flexibility such as \cite{Kreps}) and divisible commodites (auction design such as \cite{Myerson}).\footnote{We recognize that the lack of finiteness becomes an issue in propositional logic and that there are more expressive logics without this limitation. However, it is not clear to us how these may be used in the problems cited above.}
\end{enumerate}
	In conclusion, TL and GE are valuable contributions to social choice theory. We feel that automated reasoning will be a valuable addition to the tool-kit of social choice theorists.
\section*{Appendix}
\noindent We provide statements of results in social choice that have been referred to in the paper.

Let $A$ and $N=\{1,2,\cdots,n\}$ denote the set of candidates and voters, respectively.\footnote{All sets are assumed to be finite for the purposes of this article.} Each voter $i\in N$ has a ranking over the set of candidates which is a strict {\it order} $P_i$ over $A$.\footnote{$P_i$ is a strict {\it  order} over $A$ if it is a binary relation over $A$ satisfying the following properties:
\begin{itemize}
\item	Completeness : ($\forall a,b\in A$)[$aP_ib \lor bP_ia$]
\item	Reflexivity  : ($\forall a\in A$)[$aP_ia$]
\item	Transitivity : ($\forall a,b,c\in A$)[$[aP_ib, bP_ic]\implies[aP_ic]$]
\item	Antisymmetry : ($\forall a,b\in A$)[$aP_ib \implies \lnot bP_ia$].
\end{itemize}
} The set of all orders over $A$ be denoted by $\mathcal{P}$. The {\it Kendall distance} between two orders is the number of pairwise disagreements between the two orders. A binary relation $L$ over $A$ generates a {\it choice function} if ($\forall B\subset A$ , $B\neq\varnothing$)($\exists b\in B$)[$bLa$ $\forall a\in B$]. The set of all such binary relations is denoted by $\mathcal{L}$. A typical element of $\mathcal{P}^n$ is denoted by $P\equiv (P_1,\cdots,P_n)$ and is referred to as a {\it preference profile}.

	An {\it Arrovian Social Welfare Function (ASWF)} is a map $F:\mathcal{P}^n\rightarrow \mathcal{P}$. An ASWF $F$ assigns a {\it social order} $F(P)$
to every preference profile $P$.  A {\it Social Decision Function (SDF)} is a map $S:\mathcal{P}^n\rightarrow \mathcal{L}$. The axioms below pertain to ASWFs and SDFs.

\bigskip
\begin{itemize}
\item	The {\it ASWF} $F$ satisfies {\it Weak Pareto (WP)} if,	\\
	$(\forall P\in \mathcal{P}^n , \forall a,b\in A)[[aP_ib$ $\forall i\in N]\implies [a\hat{F}(P)b]]$\footnote{Given $\theta\in\mathcal{P}$, $\hat{\theta}$ denotes its {\it strict component}, i.e., $[a\hat{\theta}b]\Leftrightarrow[a\theta b, \lnot b\theta a]$ , $\forall$ $a,b\in A$.}.	\\

\item	The {\it ASWF} $F$ satisfies {\it Independence of Irrelevant Alternatives (IIA)} if,	\\
	$(\forall P,P^{\prime}\in\mathcal{P}^n$ $,$ $\forall a,b\in A)[[P\vert_{a,b}=P^{\prime}\vert_{a,b}]$\footnote{Given $\theta\in\mathcal{P}$ and $a,b\in A$, $P\vert_{a,b}$ denotes the {\it restriction} to $\{a,b\}$ of $\theta$.}$\implies [F(P)\vert_{a,b}=F(P^{\prime})\vert_{a,b}]]$.	\\

\item	The {\it ASWF} $F$ is {\it dictatorial} if,	\\
	($\exists i\in N)(\forall a,b\in A$ $,$ $P\in\mathcal{P}^n$)[$aP_ib$ $\implies$ $a\hat{F}(P)b$].\\
	voter $i$ is called a {\it dictator} in this case.	\\

\item	The {\it ASWF} $F$ is {\it anti-dictatorial} if,	\\
	($\exists i\in N)(\forall a,b\in A$ $,$ $\forall P\in\mathcal{P}^n$)[$aP_ib$ $\implies$ $b\hat{F}(P)a$].	\\

\item	The {\it ASWF} $F$ satisfies {\it Non-Imposition (NI)} if,	\\
	$(\forall a,b\in A)(\exists P\in\mathcal{P}^n)[aF(P)b]$.	\\

\item	The {\it SDF} $S$ satisfies {\it Unanimity (U)} if,	\\
	$(\forall P\in \mathcal{P}^n , \forall a,b\in A)[[aP_ib$ $\forall i\in N]\implies [a\hat{S}(P)b]]$.	\\

\item	Let $S$ be an {\it SDF}. An individual $i\in N$ is {\it decisive} if,	\\
	($\exists a_1,a_2\in A$ $,$ $a_1\neq a_2$)($\forall P\in\mathcal{P}^n$)[$a_1P_ia_2$ $\Leftrightarrow$ $a_1S(P)a_2$].	\\
	The {\it SDF} $S$ is {\it Liberal (L)} if there are at least two decisive individuals.	\\
\end{itemize}

	An ASWF satisfies WP if $F$ respects consensus, i.e. the social order ranks $a$ over $b$ whenever all voters rank $a$ over $b$. It satisifies IIA if the social ranking over any pair of candidates depends only on individual voter rankings over that pair. It is dictatorial if the social ranking coincides with that of one voter at all profiles. It is anti-dictatorial if the social ranking is the inverse of the ranking of a given voter at all profiles. It is NI if all rankings over pairs can arise as the social ranking at some profile. Unanimity is the counterpart of WP for SDFs. A voter is decisive if there exists a pair of candidates such that the voters ranking over that pair coincides with that of the social ranking at all profiles.

	Some important results in Arrovian aggregation theory are stated below:
\newtheorem*{Defn1}{Arrow's Impossibility Theorem:}
\begin{Defn1}
Suppose $\vert A\vert\geq 3$. A {\it ASWF} which satisfies {\it IIA} and {\it WP} must be dictatorial.
\end{Defn1}
\newtheorem*{Defn2}{Wilson's Theorem:}
\begin{Defn2}
Assume $\vert A\vert\geq 3$. A {\it ASWF} which satisfies {\it IIA} and {\it NI} must be null or dictatorial or anti-dictatorial.
\end{Defn2}
\newtheorem*{Defn5}{Sen's Theorem on  the Impossibility of the Paretian Liberal:}
\begin{Defn5}
Assume $\vert N\vert\geq 2$ and $\vert A\vert\geq 3$. There is no {\it SDF} satisfying {\it U} and {\it L}.
\end{Defn5}

	Strategic social choice theory is concerned with {\it Social Choice Functions (SCFs)} which are maps, $f:\mathcal{P}^n\rightarrow A$. An SCF picks the ``socially optimal'' outcome $f(P)$ at every profile $P$. In this model, a voter's order is private information. An SCF $f$ is strategy-proof if no voter has an incentive to misreport her true preferences irrespective of the report of the other voters, i.e. truth-telling is a weakly-dominant strategy for every voter. An SCF is Monotonic if it continues to pick the same candidate whenever it ``improves'' in the ranking of all voters. Finally, an SCF is dictatorial if it always picks the top-ranked candidate of a given voter at all profiles. These axioms are stated formally below:

\bigskip
\begin{itemize}
\item	The {\it SCF} $f$ satisfies {\it Monotonicity (M)} if, $\forall P,P^{\prime}\in\mathcal{P}^n,$	\\
	$[f(P)P_ib\implies f(P)P^{\prime}_ib$ $,\forall i\in N$ $,\forall b\in A]$ $\implies [f(P^{\prime})=f(P)]$. \\
\item	The {\it SCF} $f$ is {\it Strategy-Proof} (SP) if,	\\
	($\forall i\in N$ $,$ $\forall P\in\mathcal{P}^n$)($\nexists P^{\prime}_i\in\mathcal{P}$)[$f(P^{\prime}_i,P_{-i})P_if(P^{\prime}_i,P_{-i})$].\\
\item	The {\it SCF} $f$ is {\it Efficient} (EFF) if, $f(P)=a \implies (\nexists b)(\forall i\in N)[bP_ia].$	\\
\item	Let $\sigma :N\rightarrow N$ be a bijection. For all $P=(P_1,\cdots,P_n)$, $P^{\sigma}$ denotes the profile $(P_{\sigma (1)},\cdots,P_{\sigma (n)})$. The {\it SCF} $f$ is {\it Anonymous} (ANON) if, $(\forall \sigma)(\forall P)[f(P)=f(P^{\sigma})].$	\\
\item	The {\it SCF} $f$ is {\it Dictatorial} if, $(\exists i\in N)(\forall P\in\mathcal{P}^n)[f(P)=\max(P_i)]$.	\\
\end{itemize}

	Two important results in strategic social choice are stated below:
\newtheorem*{Defn4}{Gibbard-Satterthwaite Theorem}
\begin{Defn4}
Assume $\vert A\vert\geq 3$. A {\it SCF} which satisfies {\it SP} and is onto must be dictatorial.
\end{Defn4}
\newtheorem*{Defn3}{Muller-Satterthwaite Theorem}
\begin{Defn3}
Assume $\vert A\vert\geq 3$. A {\it SCF} which satisfies {\it P} and {\it M} must be dictatorial.
\end{Defn3}

	A fundamental result for domains of single-peaked preferences is \cite{Moulin} which we now describe. Let $>$ be a strict order on the set $A$. We say that $P_i$ is {\it single-peaked} if, for all $a,b \in A$, $[a<b\leq \max(P_i)] \mbox{ or }[\max(P_i)\leq b<a] \implies [bP_ia]$.

	Let ${\mathcal D}^{SP} \subset {\mathcal P}$ denote the set of all single-peaked preferences. Let $B =\{b_1, b_2, \ldots, b_{2n-1}\} \subset A$. Let $\text { median } (B) = b_j$ if $|\{b_k \ \vert \ b_k \leq b_j\}|=\frac {n}{2}$ and $|\{b_k \ \vert \ b_j \leq b_k\}|=\frac {n}{2}$. The SCF $f$ is a {\it median voter rule} if there exist $a_1, \ldots, a_{n-1}$ such that $f(P)=\text { median }\{\max(P_1), \max (P_2), \ldots, \max(P_n), a_1, \ldots, a_{n-1}\}$ for all $P \in [{\mathcal P}^{S}]^n$.
\newtheorem*{theorem}{Moulin's Theorem}
\begin{theorem}
A SCF $f:[{\mathcal D}^{SP}]^n\rightarrow A$ satisfies ANON, EFF and SP iff it is a median voter.
\end{theorem}

	We now consider the set-ranking problem considered in GE. Let $X$ be a set of alternatives with a order, $\dot{>}$ defined on it. Let $\mathcal{X}$ be the set of non-empty subsets of $X$, and let $\succsim$ be a {\it weak order} \footnote{A weak order is one which satisfies all axioms of an order except anti-symmetry.} over $\mathcal{X}$. Let $\succ$ and $\sim$ denote the asymmetric and symmetric components of $\succsim$, respectively.

	The literature on set-ranking considers the the extension of the order $\dot{>}$ over $X$ to an order $\succsim$ over $\mathcal{X}$. We briefly state some standard axioms and a basic result in this context.

\bigskip
\begin{itemize}
\item	The order $\succsim$ satisfies the {\it G\"{a}rdenfors Principle (GF)} with respect to $\dot{>}$ if,
	\begin{itemize}
	\item [(i)]	$((\forall a\in A)x\dot{>}a)\implies A\cup\{x\}\succ A$ , $\forall$ $x\in X$, $A\in\mathcal{X}$ and
	\item [(ii)]	$((\forall a\in A)x\dot{<}a)\implies A\cup\{x\}\prec A$ , $\forall$ $x\in X$, $A\in\mathcal{X}$.
	\end{itemize}
\item	The order $\succsim$ satisfies {\it Independence (IND)} with respect to $\dot{>}$ if,	\\
	$A\succ B\implies A\cup\{x\}\succsim B\cup\{x\}$ , $\forall$ $A,B\in\mathcal{X}$, $x\in X\setminus(A\cup B)$.	\\
\end{itemize}

	Suppose we add a new element to a set. According to the first part of GF, if this element is strictly better than all existing elements of the set the resulting set is strictly better than the initial one. The second part of GF says, if this element is strictly worse than all existing elements of the set the resulting set is strictly worse than the initial one. IND says the following:  If a set is strictly preferred to another, then adding an element (not contained in either set) to both does not reverse the set-ranking.

\newtheorem*{Theorem}{Kannai-Peleg Theorem}
\begin{Theorem}
Let $\vert X\vert\geq 6$. There does not exist any weak order $\succsim$ on $\mathcal{X}$ satisfying the G\"{a}rdenfors principle (GF) and Independence (IND).
\end{Theorem}

\bibliography{the_article}

\begin{thebibliography}{10}

\bibitem{Arlegi}
Ritxar Arlegi.
\newblock A note on bossert, pattanaik and xu's - choice under complete
  uncertainty: Axiomatic characterization of some decision rules.
\newblock {\em Economic Theory}, 22(1):219--225, 2003.

\bibitem{ACS}
Navin Aswal, Shurojit Chatterji, and Arunava Sen.
\newblock Dictatorial domains.
\newblock {\em Economic Theory}, 22(1):45--62, 2003.

\bibitem{Bossert}
W.~Bossert, Prashanta~K. Pattanaik, and Y.~Xu.
\newblock Choice under complete uncertainty: Axiomatic characterization of some
  decision rules.
\newblock {\em Economic Theory}, 16(2):295--312, 2000.

\bibitem{GeistP}
Christian Geist and Ulle Endriss.
\newblock Automated search for impossibility theorems in social choice theory:
  Ranking sets of objects.
\newblock {\em Journal of Artificial Intelligence Research}, 40:143--174, 2011.

\bibitem{GibbardR}
Alan~F. Gibbard.
\newblock Manipulation of voting schemes that mix voting with chance.
\newblock {\em Econometrica}, 45:665--681, 1977.

\bibitem{Kreps}
David~M. Kreps.
\newblock A representation theorem for ``preferences for flexibility''.
\newblock {\em Econometrica}, 47(3):565--578, 1979.

\bibitem{Malawski}
Marcin Malwaski and Lin Zhou.
\newblock A note on social choice theory without the pareto principle.
\newblock {\em Social Choice and Welfare}, 11(2):103--107, 1994.

\bibitem{Moulin}
Herv\'{e} Moulin.
\newblock On strategyproofness and single-peakedness.
\newblock {\em Public Choice}, 35(4):437--455, 1980.

\bibitem{Myerson}
Roger~B. Myerson.
\newblock Optimal auction design.
\newblock {\em Mathematics of Operations Research}, 6(1):58--73, 1981.

\bibitem{Papai}
Szilvia P\'{a}pai.
\newblock Strategyproof assignment by hierarchical exchange.
\newblock {\em Econometrica}, 68(6):1403--1433, 2000.

\bibitem{Sato}
Shin Sato.
\newblock Strategyproof social choice with exogenous indifference classes.
\newblock {\em Mathematical Social Sciences}, 57(1):48--57, 2009.

\bibitem{Schmeidler}
David Schmeidler and Hugo~F. Sonnenschein.
\newblock Two proofs of the gibbard-satterthwaite theorem on the possibility of
  a strategy-proof social choice function.
\newblock In Hans~W. Gottinger and Werner Leinfeller, editors, {\em Decision
  Theory and Social Ethics, Issues in Social Choice}, pages 227--234. D.
  Reidel, Dordrecht, 1978.

\bibitem{Sen}
Amartya~K. Sen.
\newblock Social choice theory.
\newblock In Kenneth Arrow and Michael Intriligator, editors, {\em Handbook of
  Mathematical Economics}, volume III, pages 1073--1181. North-Holland,
  Amsterdam, 1986.

\bibitem{ArunavaEL}
Arunava Sen.
\newblock Another direct proof of the gibbard-satterthwaite theorem.
\newblock {\em Economic Letters}, 70:381--385, 2001.

\bibitem{TangM}
Pingzhong Tang and Fangzhen Lin.
\newblock A computer-aided proof to gibbard-satterthwaite theorem.
\newblock (mimeo), 2008.

\bibitem{Tang}
Pingzhong Tang and Fangzhen Lin.
\newblock Computer-aided proofs of arrow's and other impossibility theorems.
\newblock {\em Artificial Intelligence}, 173:1041--1053, 2009.

\end{thebibliography}
\bibliographystyle{plain}

\end{document}